\documentclass{emulateapj}

\usepackage{subfigure,epsfig,url}

\begin{document}

\title{The Fragility of the Terrestrial Planets During a Giant Planet Instability}

\author{Nathan A. Kaib\altaffilmark{1,2} \& John E. Chambers\altaffilmark{2}}

\altaffiltext{1}{HL Dodge Department of Physics \& Astronomy, University of Oklahoma, Norman, OK 73019, USA}
\altaffiltext{2}{Department of Terrestrial Magnetism, Carnegie Institution for Science, 5241 Broad Branch Road, NW, Washington, DC 20015, USA}

\begin{abstract}

Many features of the outer solar system are replicated in numerical simulations if the giant planets undergo an orbital instability that ejects one or more ice giants. During this instability, Jupiter and Saturn's orbits diverge, crossing their 2:1 mean motion resonance (MMR), and this resonance-crossing can excite the terrestrial planet orbits. Using a large ensemble of simulations of this giant planet instability, we directly model the evolution of the terrestrial planet orbits during this process, paying special attention to systems that reproduce the basic features of the outer planets. In systems that retain four giant planets and finish with Jupiter and Saturn beyond their 2:1 MMR, we find at least an 85\% probability that at least one terrestrial planet is lost. Moreover, systems that manage to retain all four terrestrial planets often finish with terrestrial planet eccentricities and inclinations larger than the observed ones. There is less than a $\sim$5\% chance that the terrestrial planet orbits will have a level of excitation comparable to the observed orbits. If we factor in the probability that the outer planetary orbits are well-replicated, we find a probability of 1\% or less that the orbital architectures of the inner and outer planets are simultaneously reproduced in the same system. These small probabilities raise the prospect that the giant planet instability occurred before the terrestrial planets had formed. This scenario implies that the giant planet instability is not the source of the Late Heavy Bombardment and that terrestrial planet formation finished with the giant planets in their modern configuration.


\end{abstract}

\section{Introduction}

The standard model for planet formation predicts that the giant planets gravitationally interacted with a large number of smaller bodies (planetesimals) during and after the planets' formation \citep{helled14}. When the giant planets scatter planetesimals \citet{fernip84} demonstrated that Neptune, Uranus, and Saturn are more likely to scatter them inward than outward. Typically, these same planetesimals are ultimately ejected by Jupiter. To conserve angular momentum during this process, Jupiter must migrate inward over time while the outer three giant planets migrate outward. Building on this idea of planetesimal-induced migration, \citet{mal93} showed that Pluto's excited resonant orbit can be explained well if Neptune migrated significantly outward over the age of the solar system.  Furthermore, the Neptunian mean motion resonances sweeping through the Kuiper Belt during this migration could capture the entire resonant Kuiper Belt population \citep{mal95}. 

Meanwhile, \citet{thommes99} investigated the plausibility that Uranus and Neptune formed much closer to Jupiter and Saturn before being scattered outward by these gas giants. They found that if the primordial Kuiper Belt was much more massive and extended further inward than the modern one, dynamical friction caused by close encounters with planetesimals would recircularize Uranus' and Neptune's orbits near their present-day locations. This same process could also deplete the mass of the Kuiper Belt and explain its current excited orbital distribution. In the \citet{thommes99} work the rapid accretion of gas (and hence mass) provided the instability that led to the scattering of Uranus and Neptune. However, if the giant planets' original formation was compact enough, Saturn would initially orbit interior to its 2:1 mean motion resonance with Jupiter. As the primordial Kuiper Belt is dynamically eroded by the giant planets, planetesimal driven migration causes Jupiter and Saturn to drift apart. Consequently, they eventually cross the 2:1 MMR, which destabilizes the orbits of Uranus and Neptune, leading them to scatter off of each other, and potentially Jupiter and Saturn \citep{tsiganis05}. This scenario of giant planet evolution is now called the ``Nice Model.'' Depending on the Nice Model's initial configuration and rate of migration, the giant planet instability could be delayed for hundreds of Myrs after the solar system formed, providing a potential explanation for the Late Heavy Bombardment (LHB) in the lunar crater record \citep{gomes05, hart00}. Moreover, this giant planet instability seems to nicely explain the structure of the Kuiper belt, Jovian Trojans, and the giant planets' irregular satellites \citep{lev08, morb05, nes07, nes15a, nes15b}. 

It was quickly realized, though, that Saturn's traverse of the 2:1 MMR could alter the orbits of the terrestrial planets. During this process, Jupiter's precession rate changes, and the planet passes through secular resonances with the terrestrial planets \citep{bras09}. If the MMR-crossing takes longer than $\sim$1 Myr, there is a high probability that Venus' and Mercury's eccentricities will be excited beyond their current values. As a result, the modern terrestrial planet orbits would have a larger angular momentum deficit (AMD) than what is observed today.\footnotemark\footnotetext{Angular momentum deficit is the difference between the z-component of an orbit's angular momentum and the total angular momentum of a circular orbit with the same semimajor axis lying in the invariable plane \citep{laskar97}.} To avoid exciting the terrestrial planets' AMD, it was proposed that an ice giant directly scattered off of Jupiter, causing Jupiter to ``jump'' over the 2:1 MMR. (This scenario likely requires an extra ice giant, since ejection by Jupiter is the most likely outcome of Jovian scattering \citep{nes11, bat12}.) However, during this episode of planet-planet scattering, Jupiter's eccentricity is inevitably excited. This excitation may be communicated to the terrestrial planets via stochastic diffusion of the AMD between the terrestrial eccentricity modes and the Jovian one \citep{agnorlin12}. Indeed, even for an optimal jumping Jupiter scenario, \citet{bras13} found that the AMD of the terrestrial planets can only be consistently replicated if it was at least 70\% lower before the outer planets' instability occurred. 

Previous studies of the giant planet instability's effect on the terrestrial planets have either focused on a handful of preselected simulations or employed easily manipulated but simplified models of giant planet orbital evolution. Here we use a large suite of direct N-body simulations to statistically study the evolution of the terrestrial planets during the outer planets' instability. By performing many simulations, we can estimate the probability of the terrestrial planets' survival as well as how often their angular momentum deficit (AMD) can be kept at levels comparable to the present value. Our work is organized into the following sections: Section 2 presents the details of our numerical simulations. Following this, Section 3 evaluates the results of the simulations and the probabilities of the various outcomes. Finally, in Section 4 we discuss the ramifications of our work on current ideas of outer solar system evolution.

\section{Numerical Methods}

The Nice Model has evolved significantly since its inception, and it is now thought that 5 or 6 giant planets emerged from the solar nebula in a resonant configuration surrounded by a 20--35 M$_{\oplus}$ belt of planetesimals extending out to $\sim$30 AU. The potential sets of initial conditions is quite extensive, and \citet{nesmorb12} performed a thorough evaluation of this parameter space. They highlighted a number of giant planet resonant configurations that appear particularly promising because they are likely to yield an eccentric Jovian orbit and likely to have Jupiter and Saturn traverse their 2:1 MMR quickly. Based on their findings, we design our initial conditions using their two most preferred 5-planet configurations and their most preferred 6-planet configuration. The details of these configurations are listed in Table 1. Our initial resonant configurations are constructed via the prescription given in \citet{leepeale02} and \citet{bat10}. Once the planets are in resonance, they are surrounded by a disk of 1000 equal-mass planetesimals whose masses and semimajor axis range are set by the parameters in Table 1. Individual planetesimal semimajor axes are randomly drawn from Table 1's range to yield an $a^{-1}$ surface density profile. Planetesimals' initial eccentricities and inclinations are set below 0.01 and 1$^{\circ}$, respectively. All other orbital elements are randomly drawn from a uniform distribution. With this initial configuration in place, each simulation is integrated for 100 Myrs, which is enough time for over 95\% of our systems to pass through an orbital instability. These simulations are performed using the MERCURY hybrid integrator with a timestep of 100 days \citep{chambers99}.

\begin{table*}[htbp]
\centering
\begin{tabular}{c c c c c c c c}
\hline
Run & $N_{\rm Planet}$ & $M_{\rm Disk}$ & $\Delta$ & $r_{\rm out}$ & $a_{\rm last}$ & Resonance & Ice Giant Masses\\
Name & & (M$_{\oplus}$) & (AU) & (AU) & (AU) & Chain & (M$_{\oplus}$)\\
\hline
5GPa & 5 & 35 & 1.5 & 30 & 17.4\footnotemark & 3:2, 3:2, 3:2, 3:2 & 16, 16, 16\\
5GPb & 5 & 20 & 1.0 & 30 & 22.2& 3:2, 3:2, 2:1, 3:2 & 16, 16, 16\\
6GPa & 6 & 20 & 1.0 & 30 & 20.6 & 3:2, 4:3, 3:2, 3:2, 3:2 & 8, 8, 16, 16\\
\hline
\end{tabular}
\caption{The columns are: (1) the name of the simulation set, (2) the number of giant planets, (3) the mass of the planetesimal disk surrounding the giant planets, (4) the distance between the outermost ice giant and the planetesimal disk's inner edge, (5) the semimajor axis of the outermost ice giant, (6) the resonant configuration of the giant planets from inside to outside, and (7) the masses of the ice giants from inside to outside.}
\end{table*}

\footnotetext{For this planet configuration, \citet{nesmorb12} have the outermost planet at $a=16.1$ AU, but we find that Jupiter's final semimajor axis is better matched if this parameter is shifted to $a=17.4$ AU.}

No terrestrial planets are included in this initial phase of simulations. Modeling the orbital evolution of the terrestrial planets through a giant planet instability induced by planetesimal driven migration is a computationally expensive task. The simulation time step must be held to a small enough value to insure accurate integration of the terrestrial planets, and at least 1000 additional bodies must be included in the primordial Kuiper belt to induce giant planet migration. In addition, the instability among the giant planets can sometimes take over 100 Myrs to develop \citep{lev11}. Therefore we actually perform two integrations. The first is the integration described above that only contains the giant planets. During this 100-Myr integration, orbital elements are recorded every 10$^{5}$ years. After the simulation completes, the orbital record is searched for the first instance that any planet's eccentricity exceeds 0.1, the first time two planets' orbits cross, or the second-to-last time output before a planet is lost via ejection or collision. The simulation is restarted at whichever of these times occurs first. Upon restarting the simulation, the time step is lowered to 4.4 days (1/20 of Mercury's orbital period), and the terrestrial planets are now placed in the system. In this way, little time is spent integrating the full primordial Kuiper belt in concert with the terrestrial planets. This full system is then integrated for 200 Myrs longer. To maximize the chance that the terrestrial planets emerge intact and to minimize the total level of orbital excitation accumulated in the inner solar system, the terrestrial planets are started on their current semimajor axes with nearly circular ($e<.001$), coplanar ($i<1^{\circ}$) orbits. This approach implicitly assumes that little evolution occurs amongst the terrestrial planets before the outer solar system instability takes place. 

These simulations are designed to monitor the stability and orbital excitation of the terrestrial planets during the Nice Model. However, there are a couple different ways numerical effects could artificially produce orbital excitation or instability among the terrestrial planets. Because the MERCURY hybrid integrator employs democratic heliocentric coordinates \citep{dun98}, the accuracy of orbital integrations is degraded if the pericenter of a planet becomes significantly smaller than the smallest semimajor axis in the simulation \citep{levdun00}. This can result in non-physical drifts in semimajor axis and eccentricity of the planet and potentially lead to collisions and ejections. Such a scenario could occur in our simulations if Mercury's eccentricity is excited beyond $\sim$0.3. Empirically, we have found the that the quality of our integrations falls off significantly for these eccentricities, yet the terrestrial planets can still remain stable if Mercury's eccentricity rises above 0.3 \citep{laskar08}. For this reason, if Mercury does reach $e>0.3$ the simulation is stopped, and it is restarted from the most recent time output before this using a timestep of 1.8 days ($\sim$1/50 of Mercury's period). This smaller timestep is then used for the rest of the integration. In addition, objects are removed from these simulations if they come within 0.1 AU of the Sun, since even a 1.8-day timestep does not guarantee the accurate integration of such orbits. Planets with high enough eccentricities to attain such low pericenters are almost certainly unstable anyways.

Another artificial numerical effect may involve the planetesimals used in our simulations. Given that our planetesimal disk is comprised of bodies that can exceed half of Mercury's mass, one may wonder if these planetesimals significantly excite or destabilize the terrestrial planet orbits when they are scattered inward by Jupiter. To test the plausibility of this, we evolve the terrestrial planets with a swarm of one hundred 0.035 $M_{\oplus}$ bodies whose initial perihelia and aphelia are 0.3 AU and 8--10 AU, respectively. We then monitor how long it takes for initially circular coplanar terrestrial planets to be excited to the modern AMD via encounters with these planetesimals. On average this excitation requires $\sim$50,000 encounters between a planetesimal and a terrestrial planet (with an encounter defined as when a planetesimal comes within 1 Hill radius of a planet). In contrast, the terrestrial planets in our Nice Model simulations typically suffer only $\sim$100 such encounters. Thus, the planetesimals have a minimal effect on the final architecture of our terrestrial planet orbits.

\subsection{Scattering Experiments}

To complement our full numerical simulations of the dynamical evolution of the solar system, we also perform a batch of 1000 simple planet-planet scattering experiments. These begin with Jupiter at 5.5 AU and Saturn just beyond the 3:2 MMR with a period ratio of 1.6. In addition, one ice giant is placed 4 Hill radii beyond Saturn (a configuration that becomes quickly unstable), and the terrestrial planets are also included. All planetary orbits are initially nearly circular and coplanar ($e<.01$, $i<1^{\circ}$), and the systems are integrated for 3 Myrs with a timestep of 1.8 days. These simulations do not contain any planetesimals. To crudely compensate for the eccentricity damping provided by planetesimals, we artificially damp the eccentricities of Jupiter and Saturn with a timescale of 2.5 Myrs \citep{leepeale02}

\section{Results and Discussion}

\subsection{Evaluation Criteria}

We evaluate our simulations by whether they reproduce very basic aspects of the outer and inner solar system. In this regard, we employ three different success criteria. For the outer solar system, the simulation must finish with four surviving giant planets, and Saturn must end up between the 2:1 MMR and 3:1 MMR with Jupiter. We call this Criterion A.  If the outer solar system is reproduced at this most basic level, we then look at how many of these simulations can also reproduce the inner solar system. We have two criteria for this. First, we require that four stable inner planets survive until the end of the simulation (Criterion B). When this is the case, we integrate the entire planetary system for 1 Gyr to confirm long-term stability. Finally, for those systems with four stable inner planets, we also require the angular momentum deficit of the terrestrial planets to be equal to or less than the present value. This is our third criterion (Criterion C).

\subsection{Success Rates}

The fraction of our simulations that successfully satisfy each of our criteria are shown in Table 2. As this table shows, $\sim$10--15\% of all simulations meet Criterion A for each set of runs, confirming that the outer planets' architecture can be explained by a giant planet instability \citep{tsiganis05, nesmorb12}. However, we find that it is unlikely that both the outer and inner solar system are reproduced simultaneously. The instability of the outer solar system almost always significantly excites the orbits of the terrestrial planets. Very often this leads to the loss of at least one terrestrial planet. In fact of the 41 systems that meet our outer solar system criteria, only 3 retain all four terrestrial planets on stable orbits. Of these 3 systems, just 1 system (from simulation set 5GPb) also meets Criterion C as well. Given that each simulation set contains $\sim$100 systems, this suggests that there is only a $\sim$1\% or less chance that the giant planet instability reproduces the outer solar system's architecture while also preserving the stability and architecture of the inner planets. 

\begin{table}[htbp]
\centering
\begin{tabular}{c c c c c}
\hline
Run & $N_{\rm sim}$ & Criterion & Criteria & Criteria\\
Name & & A & A\&B & A\&B\&C \\
 & & (\%) & (\%) & (\%) \\
\hline
5GPa & 94 & 16 & 1 & 0\\
5GPb & 99 & 15 & 2 & 1\\
6GPa & 86 & 13 & 0 & 0\\
\hline
\end{tabular}
\caption{The columns are: (1) the name of the simulation set, (2) the total number of simulations run, and (3--5) the percentage of the simulations that meet our various criteria described in the text.}
\end{table}

The inner planets' orbital excitation results from a variety of mechanisms. Often, they are excited via a secular resonance between Jupiter and one or more of the terrestrial planets. It has already been shown that if Jupiter and Saturn move slowly through the period ratio range of 2.1--2.3, the frequency of the e$_{5}$ mode will temporarily match the eigenfrequencies of Venus and Mercury, and this will excite the orbital eccentricities of the terrestrial planets \citep{bras09}. One case of a terrestrial planet instability generated through this scenario is shown in Figure 1. Here Jupiter's orbital semimajor axis jumps inward at $t=400000$ years when it ejects an ice giant. As shown in panels B and E, immediately after this ejection event, the terrestrial planets' AMD is $\sim$300\% of its present-day value. However, the Jupiter-Saturn period ratio after the jump is only 2.1, and the two planets then smoothly migrate out to a period ratio of 2.3 (see panel C). During the first Myr of this migration, the eccentricities of Mercury, Venus and Earth are rapidly excited until the terrestrial planets' AMD is over 600\% of the present value. This quickly leads to a collision between Mercury and the Sun at $t=1.5$ Myrs.

\begin{figure*}
\centering
\includegraphics[scale=.65]{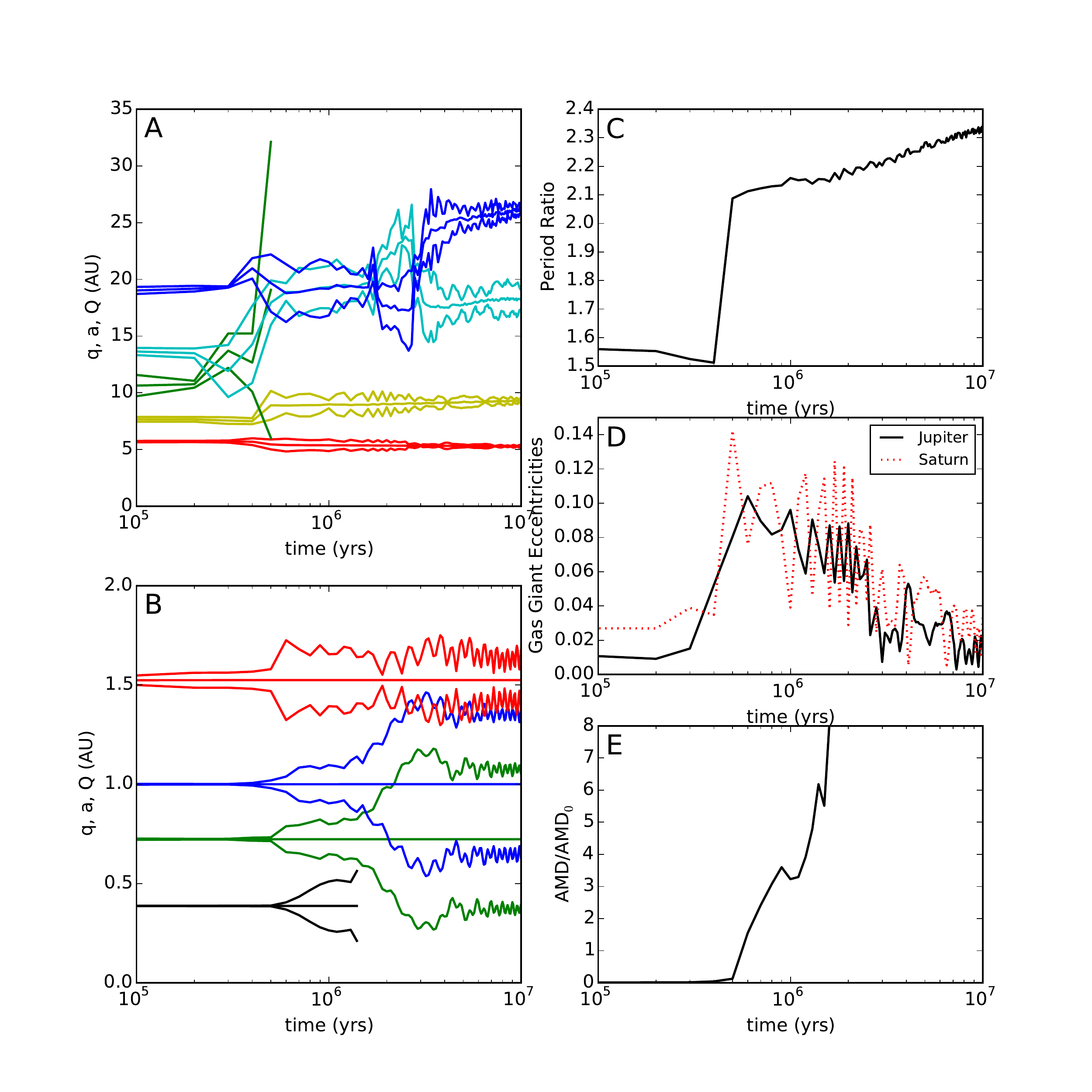}
\caption{Evolution of a system from Set 5GPa. {\bf A:} Time evolution of the pericenter, semimajor axis, and apocenter for Jupiter ({\it red}), Saturn  ({\it yellow}), an ejected ice giant ({\it green}), Uranus ({\it cyan}), and Neptune ({\it blue}). {\bf B:} Time evolution of the pericenter, semimajor axis, and apocenter for Mercury ({\it black}), Venus  ({\it green}), Earth  ({\it blue}), and Mars  ({\it red}). {\bf C:} The ratio of Saturn's orbital period to Jupiter's orbital period is plotted vs. time. {\bf D:} Jupiter's ({\it solid black}) and Saturn's ({\it red dotted}) eccentricities are plotted vs. time. {\bf E:} The angular momentum deficit of the terrestrial planets is plotted as a function of time. The simulated AMD is normalized by the AMD of the observed terrestrial planets.}\label{}
\end{figure*}

However, moving quickly through the 2.1--2.3 period ratio range does not guarantee that the terrestrial planets are protected from orbital excitation. Often the planet-planet scattering required in the Jumping Jupiter scenario can drive Jupiter's eccentricity to a value significantly higher than its current value. The $e_{5}$ eigenmode has large components in Mercury, Venus, and Earth's orbits, and significantly increasing Jupiter and Saturn's orbital eccentricity can also lead to large excitements of the terrestrial planet orbits through an exchange of AMD between the terrestrial planets and Jupiter \citep{agnorlin12}. A system from set 5GPb that illustrates this behavior is shown in Figure 2. Unlike the system in Figure 1, Jupiter and Saturn jump over the 2.1--2.3 period ratio range in well under 1 Myr, and the final period ratio settles near 2.4. In spite of this, the terrestrial planet orbits are still greatly excited during the 2:1 MMR crossing, and they emerge with an AMD that is roughly double the present value of the terrestrial planets. The source of the terrestrial planet excitation is likely related to the excitation the gas giants' eccentricities. Immediately after the resonance-crossing at $t=2.5$ Myrs, Jupiter's eccentricity is 0.077, while Saturn's is 0.20, well over three times its current value. Roughly 1.5 Myrs later, Jupiter's eccentricity hits another maximum of 0.094, and this coincides with a second rapid increase in the inner planets' AMD.  The excited terrestrial planet state is not stable, and Mercury quickly collides with the Sun. Thus, even though the requirements of the Jumping Jupiter model are met, the terrestrial planets are still destabilized. The dynamical behavior seen among our systems can be quite complex, and it is not always obvious what is the main driver of orbital excitation among the terrestrial planets. Indeed, \citet{bras13} already document cases where secular resonances between ice giants and terrestrial planets can significantly alter the inner planets' AMD. 

\begin{figure*}
\centering
\includegraphics[scale=.65]{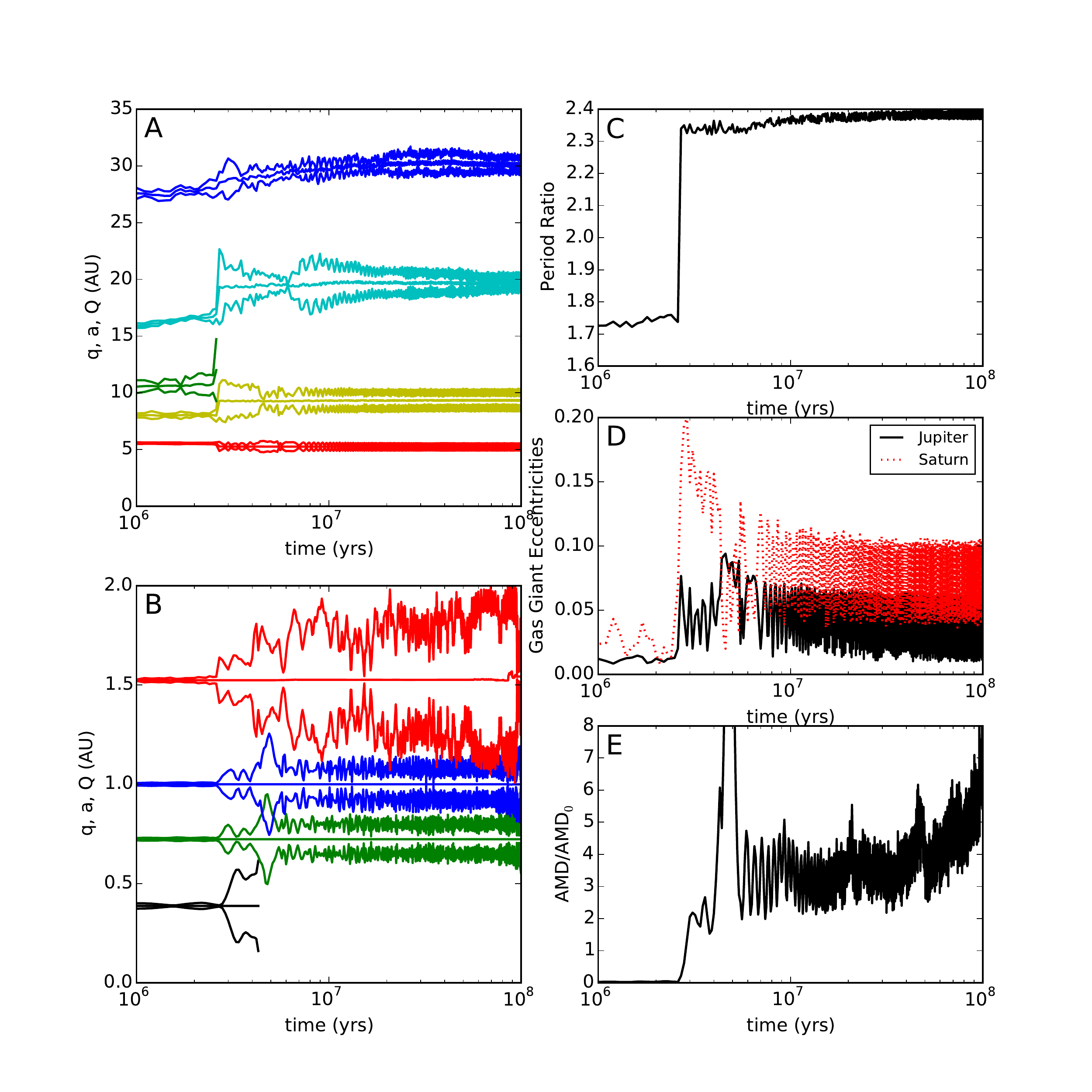}
\caption{Evolution of a system from Set 5GPa. {\bf A:} Time evolution of the pericenter, semimajor axis, and apocenter for Jupiter ({\it red}), Saturn  ({\it yellow}), an ejected ice giant ({\it green}), Uranus ({\it cyan}), and Neptune ({\it blue}). {\bf B:} Time evolution of the pericenter, semimajor axis, and apocenter for Mercury ({\it black}), Venus  ({\it green}), Earth  ({\it blue}), and Mars  ({\it red}). {\bf C:} The ratio of Saturn's orbital period to Jupiter's orbital period is plotted vs. time. {\bf D:} Jupiter's ({\it solid black}) and Saturn's ({\it red dotted}) eccentricities are plotted vs. time. {\bf E:} The angular momentum deficit of the terrestrial planets is plotted as a function of time. The simulated AMD is normalized by the AMD of the observed terrestrial planets.}\label{}
\end{figure*}

Nevertheless, one consistent aspect of our simulations is that Jupiter and Saturn typically spend time during and immediately after the outer instability with orbital eccentricities significantly greater than their current values. As illustrated in Figure 2, these states can have large effects on the terrestrial planet orbits. In Figure 3A, we look at the cumulative distribution of eccentricity maxima for Jupiter for all of our systems that meet criterion A. For our simulation sets that begin with 5 giant planets, Jupiter is typically excited to an eccentricity of at least $\sim$0.08 when the extra ice giant is ejected, and eccentricities above 0.1 occur in 1/3 of systems. In the simulations begun with 6 giant planets, the effect of planet-planet scattering is even stronger, and Jupiter typically reaches an eccentricity of 0.13. By the end of the simulations, Figure 3B shows that dynamical friction from the planetesimal population eventually damps down the e$_{55}$ component\footnotemark. \footnotetext{e$_{55}$ values are measured by integrating the giant planets in isolation for 10 Myrs and performing a frequency modulated fourier transform on the planetary orbital elements \citep{laskar99}.} In simulation set 5GPa, $\sim$85\% of the e$_{55}$ values are below the actual solar system's value, while sets 5GPb and 6GPa bracket the present-day solar system more closely. Figures 6C and 6D illustrate similar but even more extreme behavior for Saturn. In simulations begun with 5 giant planets, Saturn is typically excited to values of 0.14, while our simulations starting with 6 giant planets usually see Saturn excited to eccentricities above 0.2! Again, by the end of the simulations, Saturn is usually circularized substantially. While the final mean eccentricities seen in 5GPb systems tend to exceed Saturn's observed value, the real solar system is well-bracketed by the 5GPa and 6GPa simulation sets. All of the distributions in Figure 3 demonstrate that during the outer solar system instability Jupiter and Saturn likely have eccentricities 150--250\% of their current values when they cross their 2:1 MMR. Our simulations suggest that this has dramatic consequences for the survival of the terrestrial planets.

\begin{figure*}
\centering
\includegraphics[scale=.45]{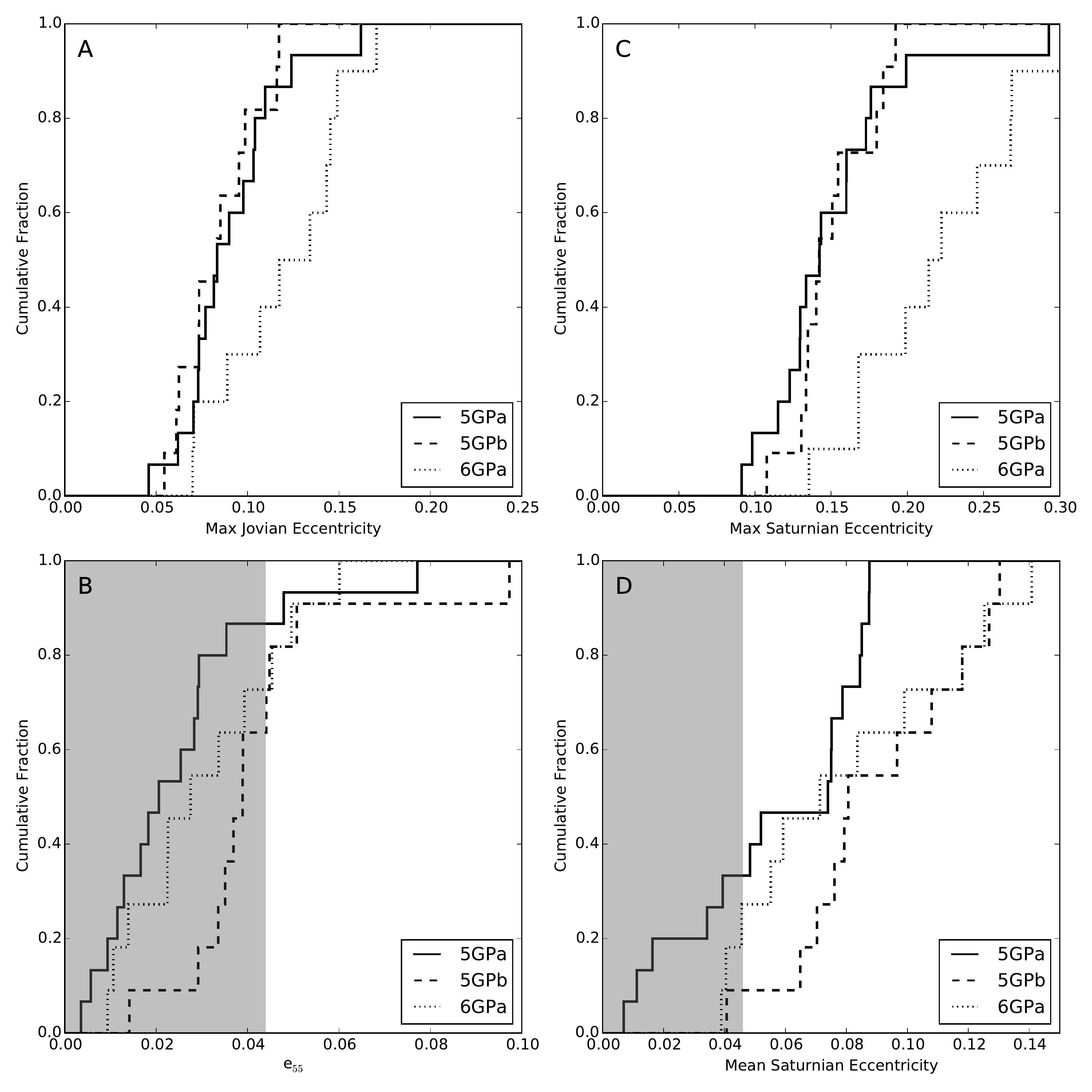}
\caption{{\bf A:} The cumulative distribution of the maximum eccentricity that Jupiter attains in each of our simulations that satisfy criterion A. {\bf B:} The cumulative distribution of e$_{55}$ amplitudes measured at the end of each simulation that satisfy criterion A. The shaded region marks e$_{55}$ values that are less than the modern value. {\bf C:} The cumulative distribution of the maximum eccentricities that Saturnian attains in each of our simulations that satisfy criterion A. {\bf D:} The cumulative distribution of the mean eccentricity that Saturn possesses at the end of each simulation that satisfies criterion A. The shaded region marks mean eccentricities that are less than the modern value. }\label{}
\end{figure*}

In Figure 4A, we look at the individual survival rates of each terrestrial planet when our simulations satisfy criterion A. We see that in simulations starting with five giant planets, Mercury is the planet most easily lost, and it survives in less than half of our systems. This is unsurprising given that it is easily excited via secular resonances. Furthermore, it is the least massive planet, so excitation of the other more massive terrestrial planet orbits can ultimately lead to Mercury's ejection. In our simulations starting with six giant planets, Mars is also easily lost. Less than 20\% of these systems retain Mars, whereas 70--80\% of systems starting with five giant planets possess Mars at the end. It appears that the large eccentricities Jupiter and Saturn reach in our 6GPa simulations are particularly disruptive to Mars. Meanwhile, Earth and Venus usually survive in all of our systems, although these rates are again lower in 6GPa systems. Some of this is by design because when two planets collide, we consider the less massive one lost, and typically this is Mercury or Mars. However, it is also partly due to the large masses of Earth and Venus. No matter which terrestrial planet is initially excited, it is likely that the end result is the loss of the lower mass Mercury or Mars rather than Earth or Venus.

\begin{figure*}
\centering
\includegraphics[scale=.6]{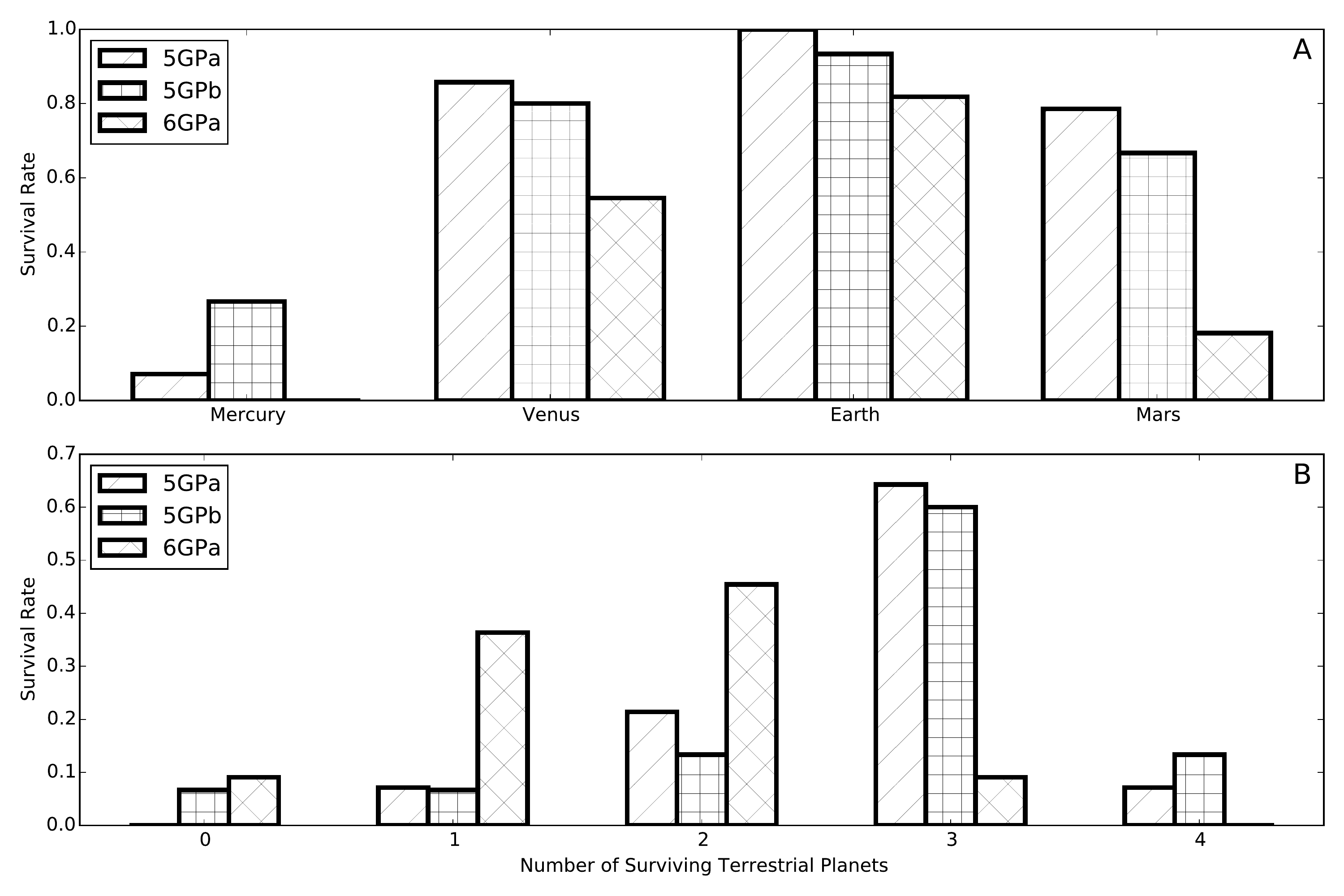}
\caption{{\bf A:} The survival rate of each terrestrial planet in systems {\it that satisfy criterion A} is shown for each of our simulation sets. {\bf B:} The distribution of the number of planets that survive in systems satisfying criterion A in each of our simulation sets.}\label{}
\end{figure*}

In many simulations, more than one terrestrial planet is lost, and in Figure 4B we show the distribution of the number of surviving planets in systems that meet criterion A for each of our simulation sets. When criterion A is met, retaining all four terrestrial planets is never the most likely scenario. Such an outcome accounts for 7, 13, and 0\% of systems in 5GPa, 5GPb, and 6GPa, respectively. For simulation sets 5GPa and 5GPb, the most common outcome is the loss of one terrestrial planet (typically Mercury). This accounts for 33\% and 47\% of systems satisfying criterion A in simulation sets 5GPa and 5GPb, respectively. Another thing borne out in Figures 4A and 4B is that the 6GPa simulations are significantly more destructive to the terrestrial planets compared to our simulations starting with five giant planets. There are no instances of a 6GPa system meeting criterion A and finishing with four terrestrial planets, and $\sim$90\% of 6GPa systems lose two or more terrestrial planets. Typically, Mercury and Mars are lost. However, even Venus and Earth only have survival chances below $\sim$80\%. The cause of this extra destruction is likely the fact that these systems must eject two giant planets and therefore undergo twice as much planet-planet scattering as our 5GPa and 5GPb simulations. Jupiter and Saturn are typically driven to more extreme eccentricities during this extended planet-planet scattering, and this often has dire consequences for the terrestrial planets. 

In addition, we find that simulation set 5GPb has the greatest chance of simultaneously reproducing both the inner and outer solar system's orbits. 1\% of our 5GPb simulations (1 system out of 99) satisfy criterion A and have four surviving terrestrial planets with AMD below today's value (although it should be noted that one other system had an AMD value of $\sim$170\% of the observed value). This is the only instance of success across all three criteria. The AMD of the lone 5GPa system meeting criteria A and B is 230\% of the observed AMD. Given the very small number of cases with four surviving terrestrial planets, we cannot definitively state that 5GPb systems yield better matches to our solar system than 5GPa. This is especially true because the giant planet instability process is highly chaotic with a variety of dynamical processes operating simultaneously, and diagnosing the exact mechanism exciting the inner planets in each run is difficult. However, 5GPa does in fact have a less massive planetesimal disk than 5GPb. A less massive disk means that when Jupiter and Saturn are evolving in semimajor axis, a smaller fraction of this evolution takes place in a smooth migration regime as opposed to the ``jumping'' regime. This minimizes the sweeping of secular resonances through the terrestrial planet region, which is a powerful driver of the excitation of terrestrial planet orbits \citep{bras09}. While simulation set 6GPa also minimizes the amount of smooth migration, the larger eccentricities that Jupiter and Saturn attain offset this effect, leading to the lowest survival probabilities for the terrestrial planets.

\subsection{Scattering Experiment Outcomes}

It is clear from the example in Figure 2 that the orbits of the terrestrial planets can be significantly altered even if Jupiter and Saturn do quickly jump across their 2:1 MMR. This orbital perturbation is typically delivered during the jumping process, which coincides with Jupiter and Saturn scattering and, typically, ejecting an ice giant. To better characterize and isolate the orbital evolution that occurs during such jumps, we consult our set of 1000 pure planet-planet scattering simulations. Although these experiments lack any planetesimals, the timescale of ice giant scattering by Jupiter and Saturn is much shorter than the eccentricity damping and semimajor axis migration timescales from planetesimals. Thus, the ice giant ejection process should proceed similarly whether planetesimals are included or not. 

In a normal full simulation, after ice giant ejection occurs the planetesimals will damp down the eccentricities of Jupiter and Saturn. In contrast, the eccentricities of Jupiter and Saturn would remain large in our scattering experiments if there are no other effects included. Since the e$_{5}$ and e$_{6}$ eigenmodes have significant amplitudes in the terrestrial orbits, this could cause us to overestimate the perturbation to the terrestrial orbits. Because of this, these simulations artificially damp the eccentricities of Jupiter and Saturn with a timescale of 2.5 Myrs \citep{leepeale02}. This is a weak enough damping to allow the scattering process (which typically occurs on a timescale below 10$^{5}$ years) to proceed unencumbered, but it is also strong enough to damp Jupiter's and Saturn's eccentricities to near zero by the end of the simulation, removing the e$_{5}$ and e$_{6}$ contributions to terrestrial eccentricities, and therefore the terrestrial AMD.

Examining scattering experiments that eject an ice giant, we find that Jupiter and Saturn ``jump'' to a period ratio between 2.3 and 2.5 in 8.7\% of runs. This is the approximate period ratio range where Saturn needs to land in order to avoid exciting secular resonances between Jupiter and the terrestrial planets and also match its observed final orbital position. For these particular cases, we find that Jupiter and Saturn pass through the 2.1--2.3 period ratio range with a median timescale of $5 \times 10^{4}$ years. This is much lower than the migration timescale upper limits of $\tau<10^{6}$ years derived in previous works \citep{bras09, minmal11}.

In addition, we see a few other trends emerge in our scattering experiments. First, there is a clear correlation between Jupiter and Saturn's final period ratio and the maximum eccentricity they attain during scattering. This trend is shown in Figure 5A. We see that as the final period ratio moves from 1.7 to 2.5, the maximum eccentricity typically attained by Jupiter increases by a factor of 2--3. A similar trend is seen for Saturn. This is expected, as a larger jump in period ratio requires a stronger interaction between the lost ice giant and Jupiter and Saturn, increasing the probability that the surviving planets will be left on less circular orbits.

\citet{agnorlin12} argue that the increased eccentricities that Jupiter and Saturn attain during planet-planet scattering will be communicated to the terrestrial planets via stochastic diffusion of AMD. Our scattering results are consistent with this process. In Figure 5B, we plot the final AMD of the terrestrial planets as a function of the final period ratio of Jupiter and Saturn. As in Figure 5A, we see a clear trend with period ratio. When Jupiter and Saturn jump to a higher period ratio they are more likely to excite the AMD of the terrestrial planets. In fact, in the preferred period ratio range of the Jumping Jupiter scenario (2.3--2.5), the terrestrial planets finish with a median AMD of 150\% of their observed value. Thus, even if secular resonances between Jupiter and the terrestrial planets are largely avoided, the terrestrial planets are typically still significantly excited.

\begin{figure}
\centering
\includegraphics[scale=.43]{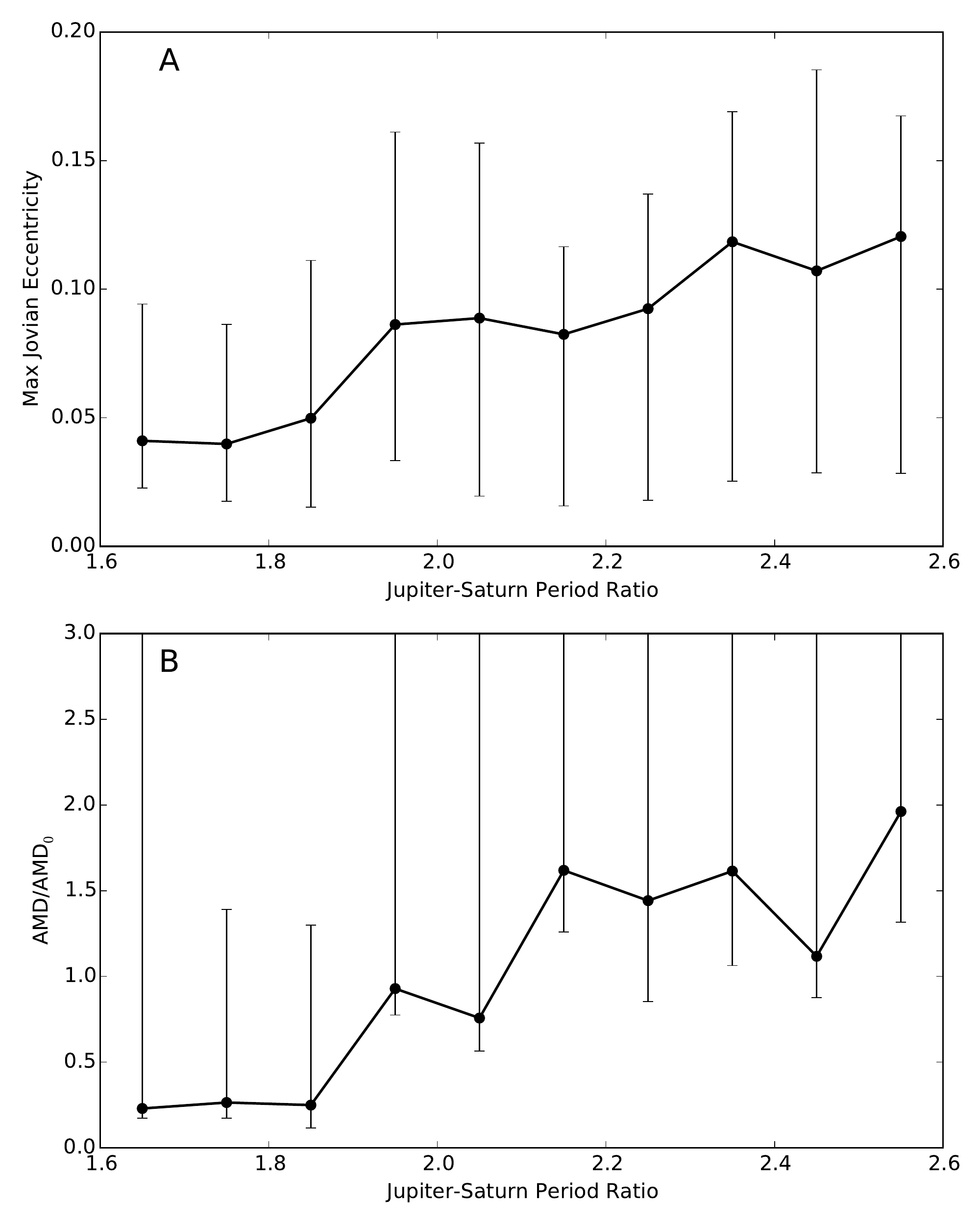}
\caption{{\bf A:} The maximum eccentricity attained by Jupiter is plotted against the final orbital period ratio of Jupiter and Saturn seen in our simple planet-planet scattering experiments. The data points mark the median values, while the error bars mark the 15.9 and 84.1 percentiles of the eccentricity distributions. {\bf B:} The final angular momentum deficit of the terrestrial planets is plotted against the final orbital period ratio of Jupiter and Saturn seen in our simple planet-planet scattering experiments. The data points mark the median values, while the error bars mark the 15.9 and 84.1 percentiles of the AMD distributions. The AMD values are normalized by the observed AMD value of the terrestrial planets.}\label{}
\end{figure}

We suspect the main culprit of this terrestrial excitation is the high eccentricity phase that Jupiter and Saturn briefly pass through during the ejection of an ice giant. To further illustrate this, we take our current solar system and perform ten different integrations of the planets with Jupiter and Saturn's eccentricity increased by a factor of two. In each simulation, the planetary inclinations are begun with their modern values, while arguments of pericenter, longitudes of ascending node, and mean anomalies are initially randomized. In these integrations, the enhanced eccentricities of Jupiter and Saturn quickly lead to instabilities among the terrestrial planets. The median timescale for the loss of the first terrestrial planet (typically Mercury) is just 8.9 Myrs. This again suggests that having Jupiter and Saturn at higher eccentricities for just a few million years can substantially modify the terrestrial planets orbits.

\section{Discussion and Conclusions}

Using three of the most promising known initial configurations of giant planets in the Nice Model, we study how the terrestrial planets behave during a giant planet instability. Our simulations demonstrate that the terrestrial planets are remarkably fragile during this process. Typically, the giant planet instability triggers another orbital instability among the terrestrial planets and at least one inner planet is lost. We find a very low fraction of simulations can simultaneously reproduce the basic orbital features of both the outer and inner planets. Depending on the initial conditions, only 2\% or less of our systems retain all four terrestrial planets and two ice giants once Jupiter and Saturn cross their 2:1 MMR and stop evolving in semimajor axis. If we do not fold the probability of reproducing the giant planets' characteristics into our analysis and restrict ourselves only to systems with four giant planets and a final Jupiter-Saturn period ratio of 2--3, we still find the probability to be 7\% or less of keeping the terrestrial planets' AMD equal to or below their observed value. We should also emphasize that our initial conditions are quite optimistic in that they begin the terrestrial planets are on nearly circular, coplanar orbits before the giant planet instability. If the terrestrial planets have significant eccentricities and inclinations before the giant planet instability, the probability of success will likely be even lower.

In fact, we only find one set of initial conditions that gives any hope of reproducing the terrestrial planet orbits. This is our simulation set 5GPb, which contained a 3:2, 3:2, 2:1, 3:2 resonant configuration for the giant planets surrounded by a 20 M$_{\oplus}$ planetesimal disk. There may be other sets of initial conditions capable of replicating the inner and outer planets' architecture, but these are currently unknown. The results of \citet{nesmorb12} suggest that any of the other configurations they explored are likely to perform worse than those evaluated here. While the \citet{nesmorb12} studies of 6-planet configurations were more open-ended, the results from our 6-planet resonant chains are very bleak for the terrestrial planets, as the violence associated with two giant planet ejections always destabilizes one or more of our inner planets.

Another consequence of our work is that the bulk of the terrestrial planets' AMD is very likely acquired during the giant planet instability. Out of 41 systems that reproduce the basic orbital features of the giant planets, there is only one system in which the terrestrial planets' AMD stays well below the modern value. The other two systems that retain all of their terrestrial planets had AMD values at least $\sim$170\% of today's value. To derive the observed terrestrial planets from these excited states requires significantly decreasing the AMD of the terrestrial planets, and there is no obvious mechanism to do this permanently. Although the terrestrial planets can exchange AMD with the giant planets, this is a reversible process, and invoking it to explain the modern terrestrial orbits requires that we currently live in a special epoch \citep{laskar08}. Additionally, dynamical friction from the planetesimals scattered during the giant planet instability can damp the terrestrial planet eccentricities, but the damping is too weak to significantly lower the AMD \citep{bras09}. One other possibility is that some of the AMD is carried off during the loss of an extra terrestrial planet, either through collision or ejection \citep{chambers07}. Figure 2E shows that the AMD can be decreased in this way. However, such a scenario is speculative and has not been demonstrated to resolve this issue.

It seems that reproducing the solar system's planetary architecture depends on events with rather low probabilities. This is due to the constraint of matching the terrestrial planets' survival and AMD. However, this constraint may not make sense if the early solar system possessed one or more additional terrestrial planets lost during the upheaval brought about by the outer solar system instability. If the original distribution of terrestrial planetary mass at all resembled the current solar system, such a planet or planets could only have existed interior to Mercury or exterior to Mars \citep{roblask01}. Furthermore, any lost planets would have likely been very low-mass. Otherwise, they would have survived instead of Mercury and/or Mars. Given that many of our terrestrial planets are lost via secular resonances with Jupiter, the additional planet would have to markedly shift the secular frequencies of one or more of the existing inner planets. Such a shift would be difficult to produce from a sub-Mercury or sub-Mars mass planet orbiting inside $\sim$0.3 AU or beyond $\sim$2 AU. In addition, if the extra planet(s) is the only object lost, it has to carry away most of the excess AMD accumulated during the instability, and this is far from guaranteed. In our systems that lose one terrestrial planet, the surviving inner planets' AMD is typically larger than the presently observed value. Considering all of this, the pre-instability secular architecture of the inner planets was unlikely to be radically different than the present one, and our overall simulation results are unlikely to be substantially affected by the inclusion of any plausible additional planets.

If we do assume that the number of terrestrial planets has not changed over the course of the solar system, the constraints their orbits provide are only relevant if the giant planet instability takes place after the terrestrial planets fully form, as discussed in previous work \citep{agnorlin12}. If the instability instead occurred early, then only certain portions of the terrestrial planet-forming disk would likely be excited and these could easily be de-excited (if necessary) via collisional damping and dynamical friction. Of course, one of the strengths of the Nice Model is that it provides a natural explanation for the Late Heavy Bombardment (LHB) seen in the lunar crater record \citep{gomes05}, and an early instability in the outer solar system reopens the puzzle of the LHB. While other potential LHB mechanisms exist, they are either not well-developed or have not held up to additional scrutiny \citep[e.g.,][]{chambers07, brasmorb11, minton15b}. However, the Nice Model LHB mechanism may be at odds with the existence of our terrestrial planets. Furthermore, in the framework of the Nice Model, it is thought that the main LHB impactor population must reside in a now-depleted inner extension of the asteroid belt near 2 AU \citep{bottke12}, but recent work indicates that the size distribution of the LHB impactors differs from main belt asteroids \citep{minton15a}. Given these recent findings and our own work, the search for additional possible LHB triggers should continue.

\section{Acknowledgements}

The majority of our computing was done using the Open Science Grid, which is supported by the National Science Foundation and the U.S. Department of Energy's Office of Science \citep{pordes07, sfili09}. We thank Kevin Walsh for helpful discussions. Finally, we thank the reviewer, David Minton, for comments and suggestions that improved the quality of this work.

\bibliographystyle{apj}
\bibliography{NiceTP}

\end{document}